# Composing Mini Oscilloscope on Embedded Systems

Brennan Romero and D.G. Perera

Department of Electrical and Computer Engineering, University of Colorado Colorado Springs

Colorado Springs, Colorado, USA

**Abstract**

In this paper, our goal is to reproduce the basic functionalities of a regular oscilloscope, using the Nuvoton NUC-140 embedded systems development platform as the front-end and display method. A custom-built daughter board connects the NUC-140 to a variety of peripherals, including two BNC scope-probe connections, an external nine-button keypad, and a calibration signal. The LCD of the NUC-140 development board serves as the waveform display. From the experimental results, it is demonstrated that our proposed system became a very competent debugging tool. It implements 90% of the features we typically use on original oscilloscopes, including: automatic, edge-triggered, and single modes; waveform visualization using vertical and horizontal scaling; probe calibration.

# Introduction

Almost all of my personal projects involve some form of electronics as a major component. These integrations have ranged from 8-bit shift register ICs wired to flip flops, 555 timer circuits, to custom-built ARM M0+ based boards. As is common in any electronics project, there are bound to be issues that arise that require some level of debug effort. My go-to tool for this kind of work is my Infinium 1.5 Ghz oscilloscope. Being able to visually see how my circuits respond in real time has cut down on hours of speculation and frustration. For ECE5330, having an oscilloscope has been instrumental in completing all of the different labs.

What makes my current oscilloscope inconvenient to use is its size and weight. My unit is from 1996, and while it is a very fast and reliable unit, it requires approximately 18.5 cubic feet of space and weighs over 30 pounds. All of my projects are small and portable, but the limited areas I can move my oscilloscope require that I bring the project to the oscilloscope instead of bringing the oscilloscope to the project. In some instances, the act of moving my project to the oscilloscope caused problems in replicating the incorrect behavior. What I need is a small, portable oscilloscope that I can bring to my projects. The speed of my current oscilloscope is overkill for most of my projects, so sacrificing speed for portability is an easy choice. The NUC140 development board presented a golden opportunity to build my own solution. With its built-in eight separate 12-bit ADC channels and plenty of unused IO, I had all of the right ingredients to make a low-speed, portable oscilloscope.

# Project Description

The project aims to reproduce the basic functionalities of a regular oscilloscope, using the NUC-140 as the front-end and display method. A custom-built daughter board will connect the NUC-140 to a variety of peripherals, including two BNC scope-probe connections, an external nine-button keypad, and a calibration signal. The two BNC connections will serve as two probe channels, allowing for standard scope probes to be attached. A waveform can then be captured on one or both of the probe channels. The nine-button keypad will serve as the system's controls, allowing for changing the trigger mode, horizontal and vertical scaling, probe calibration, and single-sample trigger. The firmware will support the following trigger modes: automatic (free running), rising/falling edge triggered, and single (user-triggered). The calibration signal provides the means of calibrating either probe channel. By attaching the corresponding probe and hitting the calibration button on the keypad, the probe channel's high and low values will be recorded to more accurately represent signal magnitudes. Two buttons located above the BNC connections will allow for enabling and disabling the different probe channels. A 3x8 pin header with jumpers will be used to tie ADC channels to a specific probe channel.

The LCD of the development board serves as the waveform display. Above the waveform, a signal line of text will indicate the trigger mode, a representation of enabled channels, the ADC_N value, and the vertical scaling factor. Below the display, the four red LEDs will indicate the current acquisition status of one of Armed, Triggered, Complete, or Displayed.

# Hardware Design

The vast majority of the hardware design was dedicated to designing, laying out, and assembling the custom daughter board for this project. The choice of building a daughter board was driven by two factors: reliable connections and breadboard-incompatible connectors. The project requires over 17 external connections to the NUC-140; from experience, having one or multiple unstable connections is a huge time sink. Taking the extra time and resources to build a PCB to make those connections as stable as possible contributes to an overall easier development cycle. This is especially important when measuring signals, as any distortion that occurs in the front end will affect the displayed waveform. The other factor that made the PCB a required component was the BNC connectors. Standard oscilloscope probes use BNC connectors as their design allows for sturdy yet easy probe insertion. However, these connectors do not share the same 2.54 mm pitch of a breadboard, nor are the size and shape of the pins compatible. While it might have been possible to solder breadboard-compatible jumper wires to BNC plugs, this action would violate the first reason stated for using a PCB.

The need for an external keypad on this daughter board is the result of the GPIO pins associated with channels 0 to 6 of the ADC being attached to the development board's nine-button keypad. The keypad is built like a keyboard matrix, where detection of a pressed key is done by sweeping a signal across rows of buttons and seeing which of the input pins receives it. Since all of the ADC channels need to be strictly inputs, this keypad could not be used.

From previous projects, I already had buttons, resistors, and 2.54 mm pin headers necessary for constructing the keypad, channel enable buttons, and headers. What I lacked for this project were BNC connectors and jumpers required to build the ADC to BNC connections. To obtain these components, I used Digikey as my supplier. I ended up using the CONBNC001 female BNC connector from TE Connectivity Linx and 2.54mm 2 x 1 pin jumpers from Sullin Connector Solutions to fulfill these requirements. Knowing which components I was going to use helped me to choose the correct footprints when doing board layout. Having a completed part list, I proceeded to design a schematic and layout for the daughter board.

Above is the full schematic used to lay out the daughter board.

The final board layout of the daughter board.

To design my board, I used KiCAD, as it is the software package I have the most experience using. The board design ended up being one of the simplest parts of the project as a whole. The design for the nine-button keypad mirrors the one on the development board to use as few pins as possible. All of the pin headers are the standard 2.54 mm pitch, just like the pin headers on the NUC-140 board. Using the same type of pin header allowed for using standard Dupont jumper wires to connect to the NUC-140. Routing the signals was relatively simple, given the low number of signals and proper planning in pin assignment. While most of the circuitry is standard, the one important feature of this board is the ADC channel selection header:

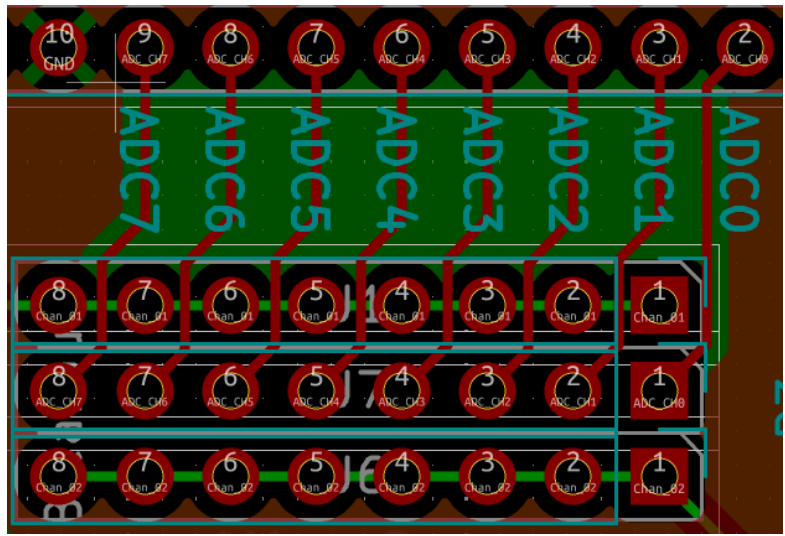

All of the GPIO pins assigned to ADC channels are routed to the middle row of pins. ADC channels can then be assigned to probe channel 0 or probe channel 1 by jumping from the middle row to the top or bottom row, respectively. For instance, to attach GPA_0 to probe channel one, you would install a jumper for GPA_0 of the middle row to the row below.

After completing the board design and a quick board review from an industry expert, I sent the board files to OshPark to have the PCBs fabricated. Upon receiving samples of my board, I soldered all of the components onto a single sample. Once fully assembled, I connected the daughter board to the NUC-140 according to the pinout, allowing me to move on to firmware development.

## Software Design

The software design took up the vast majority of the development time of this project. The feature set required for building an oscilloscope is fairly large, making use of a sizeable chunk of the NUC-140 feature set. This was also compounded by the fact that portions of the development board are under-documented, requiring reverse engineering efforts to complete development.

I started development for the firmware by exploring how to draw arbitrary graphics to the LCD display. The included driver could draw text onto the screen using a preset ASCII to graphics array, but drawing waveforms requires being able to plot individual pixels. By comparing the SPI data sent to the display controller, I was able to identify the controller as being protocol compatible with ST7565R found in a comparable display. By cross-referencing the included driver with the ST7565R manual, I was able to reverse engineer what the original driver was doing to draw to the display. Using this information, I refactored and re-wrote parts of the driver to better clarify what the driver was doing. This mainly entailed using #define's to label the different controller commands (such as labeling 0xEB as CMD_BOOST_RATIO) and writing a macro to condense the command to SPI interface dataflow.

It was during this time that I learned how to write a byte to the LCD's display RAM, allowing me to set any pixel on the display as lit or unlit. To the controller, the display is 132 pixels wide by 8 columns of 8 pixels. However, the panel is only 128 pixels wide by 64 pixels long. The 128 pixels of the display map to entries 2 to 129, with entries 0 to 1 and 130 to 131 remaining unmapped. The display RAM is contiguous in column-major order. Writing to the display is done by sending an entire 8-pixel row as a byte to the display. Each bit of this byte represents one pixel of the display along the 64 long axis. To which of the eight columns and which row the data is written is based on the value of the address register internal to the controller. Every write causes this value to increment by one row along the column in column-major order. This start address can be set by sending the Set PA (pixel address) and Set CA (column address) upper/lower bits commands to the display. In the old LCD driver, this was done by calling the SetPACA function, passing the pixel address and column address as two unsigned 8-bit values.

The layout of the display RAM and the auto-incrementing display address were the leading factors for how I implemented drawing waveforms to the display. On the NUC-140 side, I would create a framebuffer that replicates the RAM layout of the display. This was done by defining a union type that allows for column-major 2D RAM access or fully contiguous access:

```
typedef union lcd_display_u
{
    uint8_t u8_col[LCD_COLUMNS][LCD_WIDTH_LOGICAL];
    uint8_t u8_data[LCD_TOTAL_DATA];
}lcd_display_t;
```

Where LCD_COLUMNS is eight (for the eight columns of eight pixels), LCD_WIDTH_LOGICAL is 132 (128 + 4 unmaped rows), and LCD_TOTAL_DATA is LCD_COLUMNS times LCD_WIDTH_LOGICAL. To handle plotting individual pixels to an instance of this frame buffer, the following macro, PLOT_AT, was used:

```
#define MOD_LCD_COLUMN(y) ((y) & (LCD_COLUMNS - 1))
```

```
#define PLOT_AT(buff, x, y) { \
    uint8_t y_pixel = 1 << ((LCD_COLUMNS - 1) - (MOD_LCD_COLUMN(y))); \
    uint8_t col = (LCD_COLUMNS - 1) - MOD_LCD_COLUMN((y) >> 3); \
    uint8_t x_pixel = (LCD_WIDTH - 1) - (x) + 2; \
    (buff).u8_col[col][x_pixel] |= y_pixel; \
}
```

The addressing scheme here works similarly to how CPU cache is implemented. This macro determines which of the eight columns the pixel is a part of, along with which pixel within the column it lies. The macro also determines which of the rows of eight pixels along the x-axis the pixel belongs to. The rotation of the display (due to the dev-board mounting the panel upside down) is accounted for by subtracting each pixel from the total length of the axis. Once the column, y pixel, and x pixel address is known, the u8_col member of the framebuffer union is accessed at the specified column and x-pixel row. The y-pixel is then ORed into this byte. To draw this framebuffer to the display, the Pixel and Column address of the display is set to 0,0. Then, using the u8_data member to walk along the framebuffer contiguously, the display is written one byte at a time. The original method of printing ASCII text was backported to draw the framebuffer. To speed up transfers, the LCD driver rewrites to increase the SPI speed by lowering the clock divider value from three to two.

The next step of development was creating code to use the new external nine-button keypad. To accomplish this task, I wrote three functions: init_ext_keypad(), read_keypad(), and press_any_key(). init_ext_keypad() sets GPE pins 2,4, and 6 as inputs and GPE 1,3, and 5 as outputs, forming the scan and receive pins, respectively. As the name implies, read_keypad() determines which of the nine keys is being pressed, returning 0 if none are pressed. At the start of the function, read_keypad() temporarily sets GPE 2/4/6 to output and drives a low to each before setting the pins back to input mode. This is done to bleed any charge left on the pins from a previous keypad scan. Once these signals are cleared, the keypad is scanned by driving each row pin high and seeing if one of the input pins goes high as a result. The press_any_key routine allows the caller to wait for a button press by waiting for the keypad to be released, followed by a button push.

The calibration signal is generated by using the PWM subsystem of the NUC-140. To avoid conflicting with other pins, PWM channel 6 on GPE_0 was chosen to provide this signal. Setting up this channel involved:
- Setting channel 6/7's clock to the 22 Mhz internal oscillator.
- The prescaler to 1/2
- The clock divider to 1.
- CNR to MAX_FREQ, and CMR to MAX_FREQ/2.
- Signal inversion to off.

- Enabling the POE.
- Disabling capture.
- Enabling channel 6.
- Setting GPE pin 0 to PWM mode.

MAX_FREQ is a value set at the top of the main.c file, responsible for setting the frequency of the PWM based on the number of cycles of the 22 Mhz clock. By setting CMR to MAX_FREQ/2, a 50% duty cycle on the signal would be achieved. The value of the CNR register in the final version was set to 2200, which would produce a 2.5 Khz clock signal. The exact frequency of this signal was verified by attaching a real oscilloscope to the calibration pin of the daughter board:

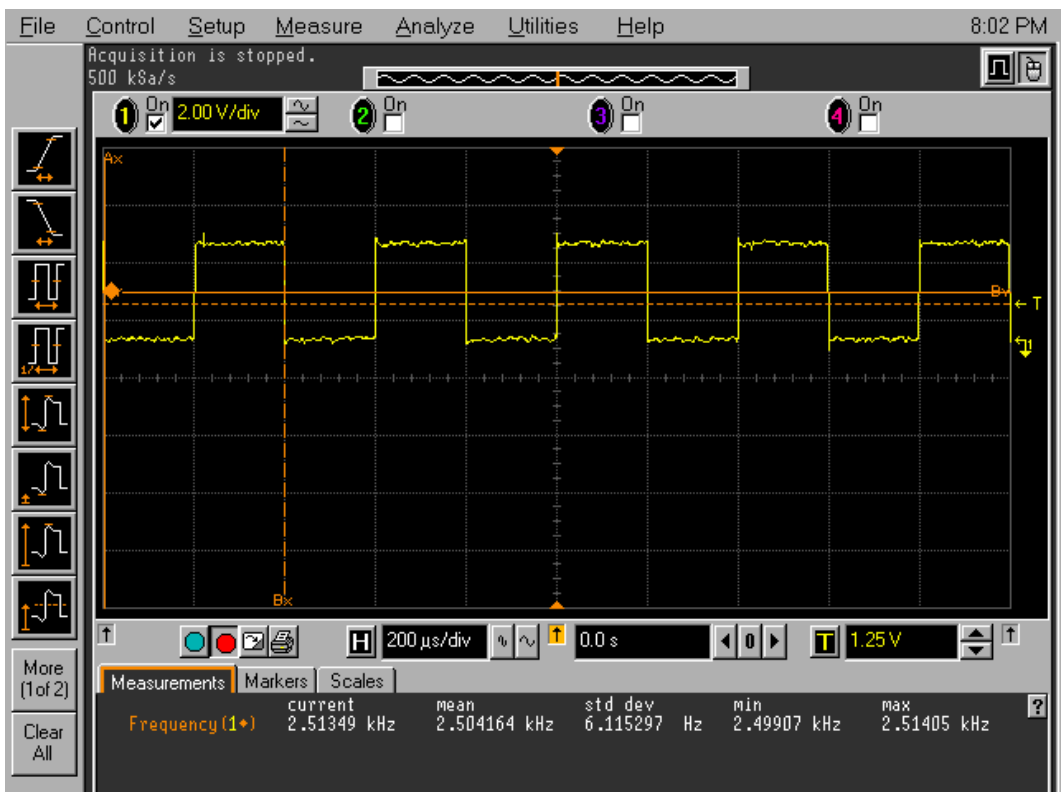

The target frequency of 2.5 Khz was chosen during testing as it provided a clear image when measured by the NUC-140.

One of the most important parts of the software design was the ADC setup. I chose to directly set the ADC registers to allow fine control over how the ADC functioned. For this project, this meant setting up and enabling all eight channels of the ADC in a continuous-scan, non-differential mode. To manage copying data out of the ADC result registers, I enabled ADC interrupts, using ADC_IRQHandler as the ISR for this event. The clock for this channel would be the 22 Mhz internal clock. To allow for horizontal scaling, the value of the clock divider ADC_N is set to a range of as small as three to as large as ten. The range allows for capturing a wide variety of signals while allowing enough time to copy data out of the ADC result registers. Not included in the ADC setup is starting the ADC by writing 1 to the ADST field of the ADCR register. This is because the start of ADC sample collection needs to occur based on the trigger mode of the system.

As mentioned previously, the ADC_ISRHandler function's job is to copy data from each channel's result register.

To make the waveform display easier, these values are copied into an array 128 * 2 entries long, called channel_data. The 128 corresponds to the number of pixels of the display, while the *2 allows for two waveforms to be stored when in dual probe-channel mode. Since one cycle of the ADC only produces eight values, a global variable within a larger system struct called buffer_idx is used to keep track of the total number of collected samples. Each time the ADC ISR is fired, it immediately clears the ADF flag to prevent the ISR from significantly delaying obtaining the next samples. Afterwards, it begins copying data into the appropriate section of the channel_data array, using the buffer_idx value to start at the correct location. Note that the full 32-bit register, including the result bits, is copied. Extracting the 12-bit field of the register requires more instructions and, therefore, more time. A separate post-processing step is performed after acquisition to extract this field once timing is no longer a concern.

Since the CPU is running faster than the ADC clock, the value of the ADC registers doesn't get a chance to change before the value is copied out. Once the copying is completed, the buffer_idx value is increased by eight for the next time the function is run. When the buffer_idx has obtained a sample count greater than the value adc_required_samples in the system struct, the ISR disables the ADC by clearing the ADSR flag. This function also clears the buffer_idx for the next waveform acquisition. One thing of note in the above function is the use of the GCC builtin function __builtin_expect(). As will be discussed later, the amount of time this function takes is critical to making sure waveforms are sampled at a uniform rate. Using __builtin_expect(), I hinted to the compiler that the code inside the if statement is unlikely to be run most of the time. This allows the compiler to perform better branch prediction, which in turn allows the ISR to prepare to exit earlier.

Triggering on rising and falling edge triggers is done by utilizing the NUC-140's built-in GPIO interrupt event. This works by setting up GPA_0 (ADC channel 0) to be able to cause the GPAB interrupt. Setup for this functionality is done by the function init_gpio(). The function first sets up interrupt debouncing using the 22 Mhz clock with a timer of $2^5$ timer ticks. While the function does enable the GPAB interrupt in the NVIC, it does not enable GPA_0 as an interrupt source. Instead, the function capture_samples() will enable the interrupt (along with the appropriate trigger type) when the system trigger mode is appropriate. If the interrupt is enabled and fired, the GPAB ISR will disable GPA_0 from triggering interrupts, clear the interrupt source, and enable the ADC.

The ISR is required to disable GPA_0 from firing future interrupts to prevent unwanted re-triggers of the ADC.

The main() function of the firmware package is responsible for calling each subsystem initialization function (such as init_keypad), clearing out the channel_data and lcd framebuffer arrays, as well as setting default values for the global system state. The global system values, alluded to previously, are stored within a global instance of the struct system_state_t called sys_state. sys_state contains values for the current trigger mode (auto, rising/falling triggered, etc), the current status of the acquisition collection (collect_state), calibration values for each probe channel, and any other values needed by the system.

Within an infinite loop, the main function first checks to see if a key has been pressed by calling read_keypad(). The exact function of each key will be detailed in the section on software/hardware integration. The main function is also responsible for reading the probe channel enable buttons and toggling them on and off. After checking all the different buttons, the system clears the top two columns of the LCD framebuffer and draws the current system status line. This status line includes the current trigger mode (A = Auto, TF/TR triggered rising/falling, Single), the current ADC_N value, which probe channels are enabled (Y for on, N for off), and the vertical scaling factor. When a full waveform has been captured, the main function will call a function called plot_data() to draw the data to the unused portion of the display. plot_data() works by taking in a pointer to a 128-entry array of ADC data, and performing scaling and translation on each value to correctly draw the waveform as a series of pixels. Linear interpolation is used to connect the dots between two side-by-side datapoints. When two probe channels are enabled at the same time, this function is called twice to plot one waveform on top of another. If the waveform has not yet been captured, the section of the framebuffer where the waveform resides is not cleared. This allows the user to continue to view a waveform while a new one is being acquired.

The function capture_samples() is responsible for starting and monitoring waveform acquisition. It is called by main within the main while loop to check and/or update the status of waveform acquisition. The exact behavior of this function depends on the value of the collect_state member of the system struct. This variable is shared between this routine and the ADC ISR. This variable has a value of one of COL_STATE_ARMED, COL_STATE_TRIGGERED, COL_STATE_DONE, or COL_STATE_DISPLAY.

When collect_state is set to COL_STATE_ARMED, the function will perform the setup required to begin waveform acquisition. If the current trigger mode of the system is either

MODE_TRIGGERED_FALLING or MODE_TRIGGERED_RISING, the function will set GPA_0's trigger method to match the requested mode, followed by enabling interrupts from GPA_0. In any other mode, the function will instead just enable the ADC by writing a 1 to the ADST flag. Upon completion, the function moves from COL_STATE_ARMED to COL_STATE_TRIGGERED. The function performs no operation while in this mode, as it is waiting for the ADC ISR to fill the channel_data array. When the ADC ISR reaches the target sample amount, the ISR will move the state from COL_STATE_TRIGGERED to COL_STATE_DONE. When capture_samples() sees this state, it will call the postproc_samples() function and then transition to the COL_STATE_DISPLAY state. This signals to the main loop to clear out the old waveform plot and plot the new data. The job of the postproc_samples() function is to extract the 12-bit result from each ADC sample. In dual-probe-channel setups, postproc_samples() is also responsible for de-interleaving the data, as a dual-probe setup requires every probe channel be attached to every other ADC channel. The main loop visualizes which state is active by turning one of the NUC-140's onboard red LEDS based on the current state.

The biggest problem encountered during development came in the form of ISR timing issues. Even though the ISR routine is relatively small, at ADC_N values less than three, the ADC collects data faster than the ISR can copy it out. As a result, values within the result registers can change during the copy operation, which causes captures of the calibration signal to become corrupted:

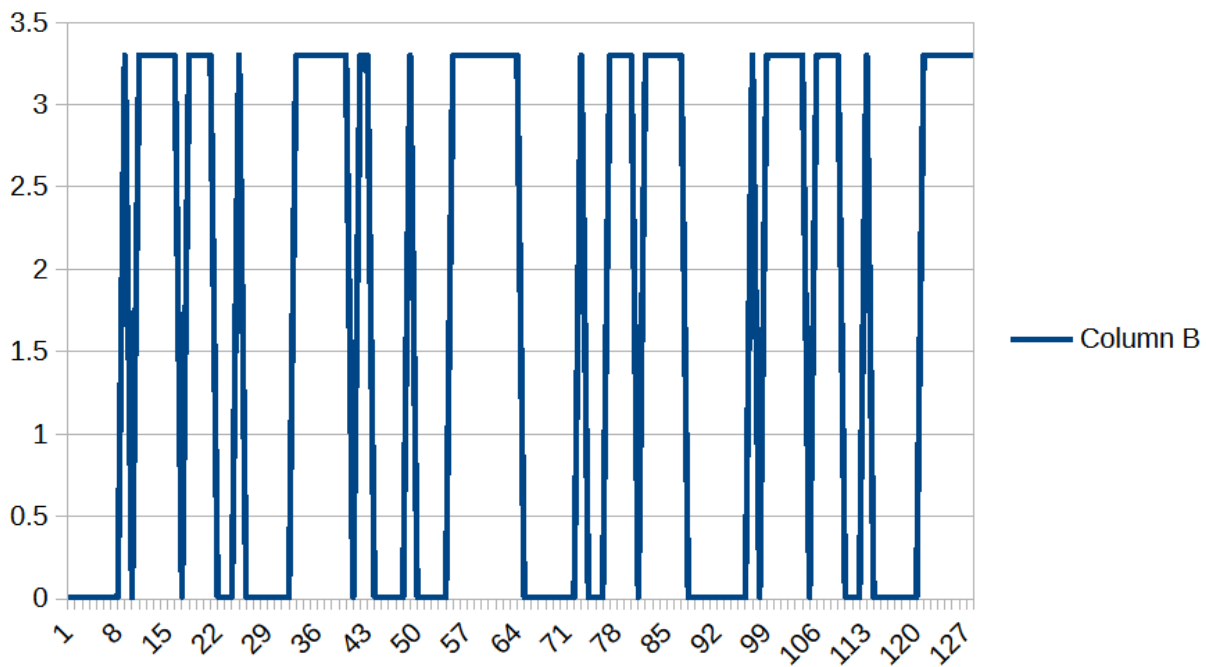

The above waveform should be a square wave. Due to timing race conditions, the waveform's data has been corrupted.

The fix for this was to limit the minimum ADC_N value to three. At a value of three, the ISR can copy data out quicker than the ADC can modify it, preventing the waveform from becoming corrupted.

## Software-Hardware Integration

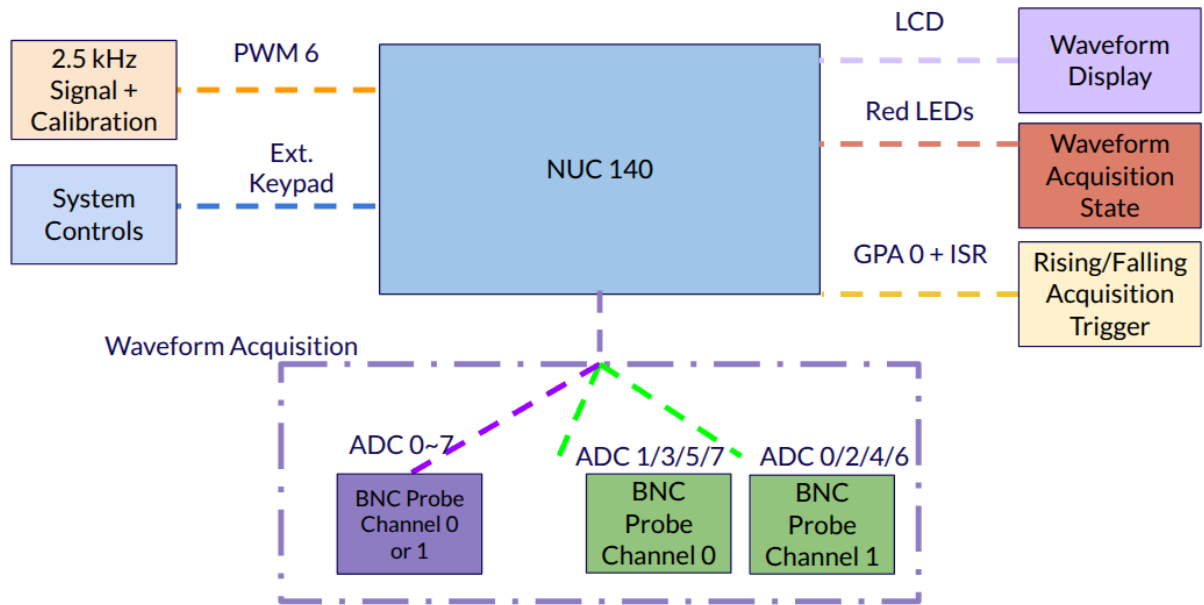

A complete system diagram showing the connections between the NUC-140's hardware and its firmware functionality.

Being a device that measures and displays the response of external circuits, the hardware and firmware are tightly coupled. Hardware triggers firmware events, and firmware controls and tunes the hardware. One of the largest examples of this coupling is found in the setup of single or dual-probe channel setups. When using a single probe channel, the software requires that the ADC-to-probe channel jumpers be set to tie all ADC channels to the correct probe channel. Dual probe-channel setup is slightly more complicated. There are only eight ADC channels, and it is mandatory to subdivide them into two groups. Using two probes tied to the same set of ADC channels can cause the signal from one probe to affect the signal from the other. The ADC subsystem of the NUC-140 is unaware of the existence of the two separate probes, and will continue to sample in ascending channel order. This is akin to the ADC acting as if the system is in single-probe mode. It is up to the firmware to post-process the collected samples to differentiate between the two separate probe-channel measurements. To ensure correct separation, the firmware and the hardware need to be in "agreement" over which ADC channel is tied to which probe channel. To minimize the largest time delta between sampling two separate signals, the firmware is set up to assume that probe channel 0 uses every even ADC channel, with probe channel 1 using every odd ADC channel. For the firmware's assumption to be correct, the jumpers on the daughter board must match the expected configuration. When the hardware and the firmware are in agreement, dual waveform capture functions correctly and as intended:

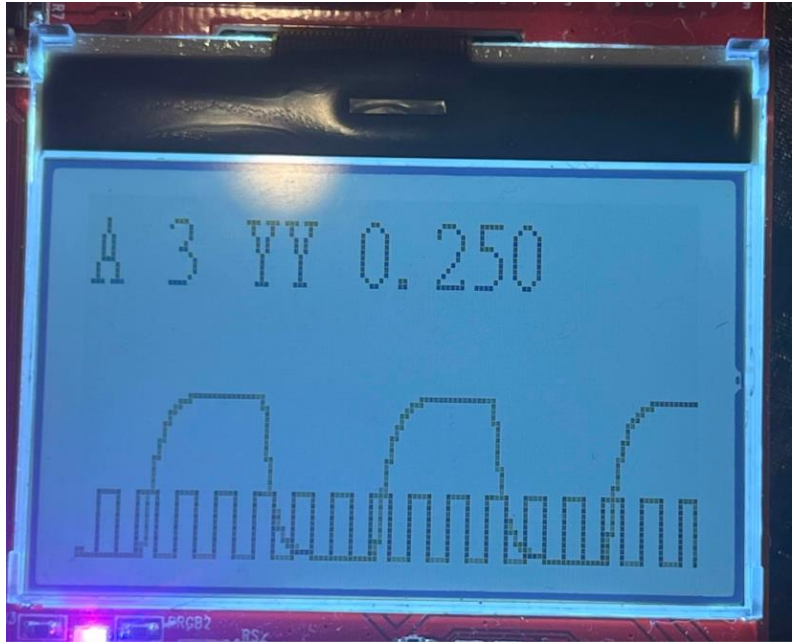

Probe channel 0 measuring a RC signal, while probe channel 1 measures a clock signal generated every ADC ISR event. Every transition of this signal identifies when all eight ADC samples have been acquired.

In the opposite direction, firmware takes on the role of calibrating each probe channel to account for the probe effect on the circuit. This is most notably required when a probe channel is attached to ADC channel 7. That specific channel causes signal distortion as the GPIO pin is attached to the wiper of a potentiometer on the NUC-140 development board. To compensate for this voltage offset, the firmware features a calibration mode using the dedicated calibration signal. Calibration for probe channels 0 and 1 is triggered using the 7 and 9 keys on the external keypad. A brief message will appear asking the user to connect the probe to the daughter board's calibration pin. After connecting the appropriate channel and pressing a button to continue, the firmware will switch the calibration pin back to GPIO mode. Firmware will then drive a high signal and trigger, and poll capture_samples() to manually perform a waveform acquisition. The average of all 128 samples will be used to determine the "high" value of the probe. Driving a low signal and repeating the process will find the "low" value of the probe. After calibration is completed, the calibration pin is returned to PWM mode, and control is returned to the user. Properly calibrated probe channels allow signals to be drawn with the correct relative amplitude.

## Testing Methodology

Verifying work was done progressively as the firmware gained features. As arbitrary graphics on the LCD came first, it was the first feature verified. This was done by plotting pixels at each corner of the display to verify the PLOT_AT macro, followed by verifying the ASCII to framebuffer code would correctly draw the text at the right location. Verifying the frequency of the calibration signal was performed by measuring the signal with a real oscilloscope. Checking that the external keypad worked correctly was done by manually stepping the probe signal and verifying each button would drive a positive voltage to its respective scan pin.

NUC-140's UART served as the main method of verifying the correctness of measurements. Once the system could support capturing waveforms, instead of displaying to the LCD, I would output the captured values in CSV over UART. Capturing these values on a PC allowed me to use LibreOffice Calc (A free version of Excel) to quickly plot the signal. By simultaneously capturing the same waveform using a real oscilloscope, I could check that my system was correctly measuring waveforms. My go-to test for this was measuring the calibration signal generated by the board using one of the probe channels:

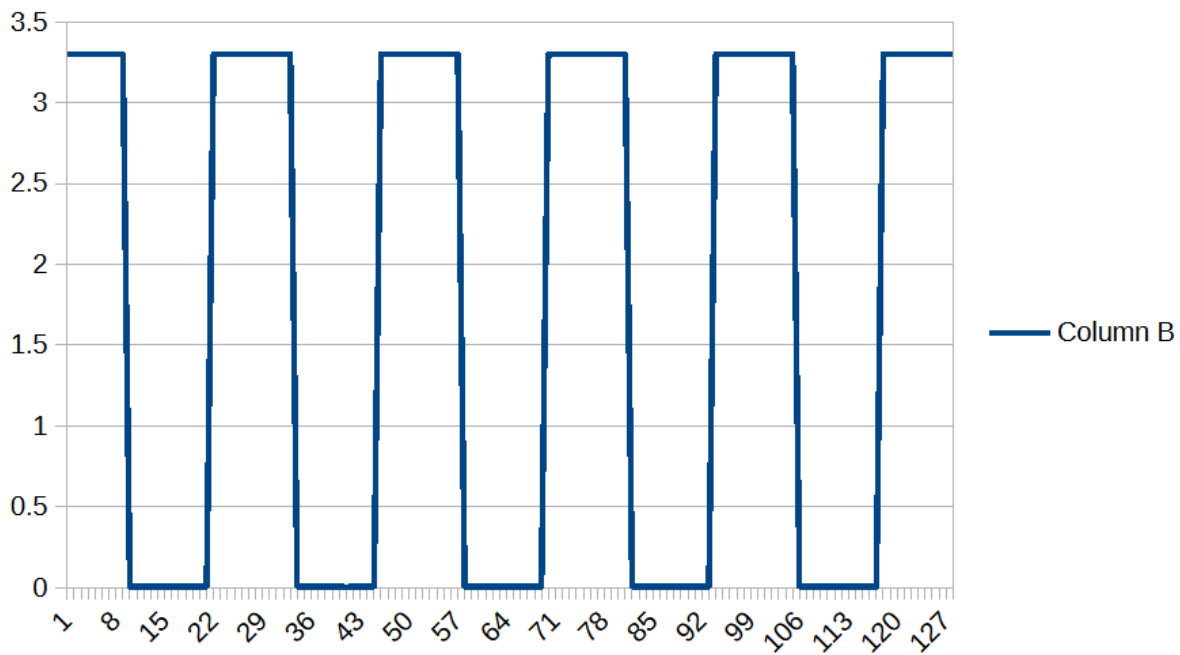

The 2.5 Khz signal, plotted from the data received via the UART.

This method of testing the calibration signal is what allowed me to debug the waveform corruption issue and identify the maximum speed at which the ADC could be run.

Once I verified the system was correctly measuring waveforms, I added support for plotting the waveforms on the display. Using the same calibration signal as the source, I could verify that the plotting routine was correctly drawing the above square wave. At this time, I used the external keypad to ensure that the ADC_N and vertical scaling adjustments worked as intended. Once I knew the waveform was being plotted correctly, I added the calibration routine, validating its correctness by having the LCD output the high and low values after completion.

Given that they were already used in previous measurement tests, the automatic and single-shot trigger modes were already known to be working. Validation of the rising and falling edge triggers was completed by enabling the respective modes and measuring the calibration signal. These modes should prevent the signal from migrating due to starting conversion at the same time (the edge of a stable clock). Both modes were found to show a steady signal, with the waveform not migrating over time. It was at this time that the debounce time of $2^5$ clock cycles was optimal for stable edge detection.

The final tests involved measuring more "exotic" signals. Up until this point, the only signals being measured were square waves with a 0 to 3.3v (Vref) swing. To ensure the system could also measure AC signals, I also measured the output of an EICO 377 audio generator. It has the capability of producing a range of sine waves, ranging from 20 to 20,000 Hz. Measuring the output and adjusting the display scaling allowed for the sine wave to be accurately drawn to the LCD display:

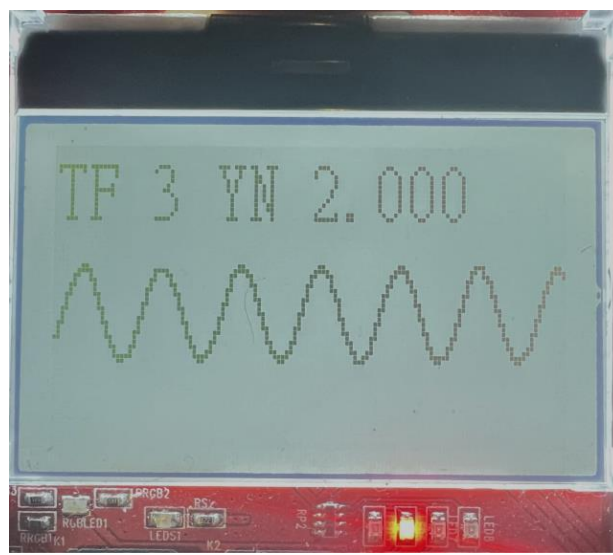

All of the tests up until this point proved that the system was working as intended. However, there was one aspect of the system that testing proved did not function correctly. As stated as a part of the project goals, this system was intended be used with regular oscilloscope probes. This was supposed to allow for higher-resolution waveform capture by taking advantage of the probes attenuation. It was revealed during testing that two factors would prevent the usage of probes with any attenuation greater than 1:1. The first was the potentiometer attached to ADC channel 7. When connected to the same channel with a 5:1 scope probe, the potentiometer would overpower the signal from the probe.

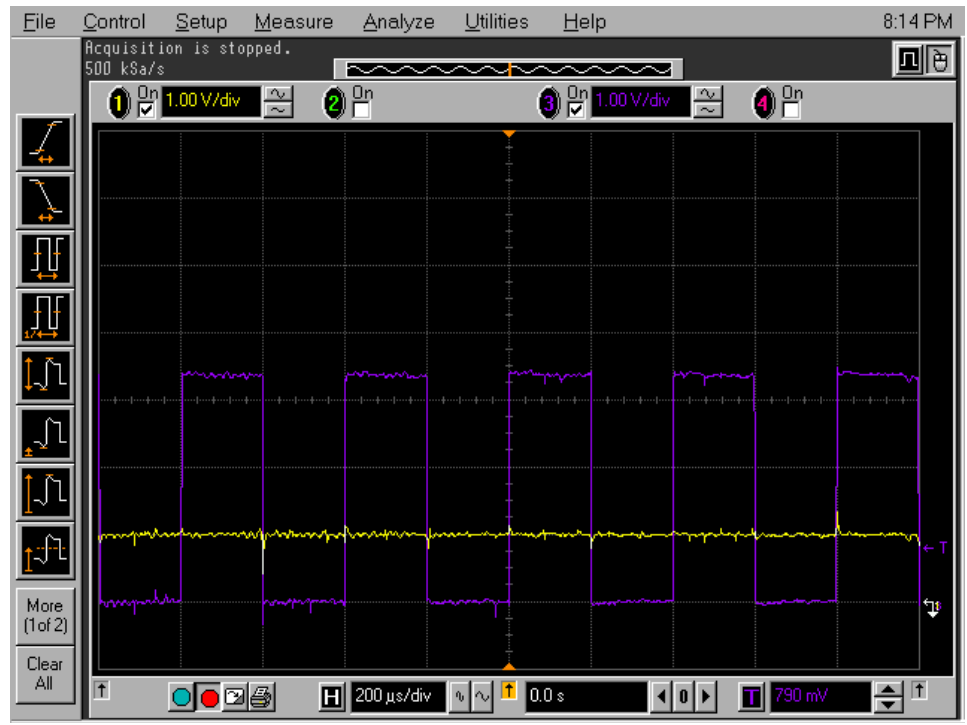

The purple line represents the signal measured at the calibration pin. The yellow line represents the signal after passing through the 5:1 scope probe connected to the daughter board. The potentiometer's influence causes the signal to remain flat.

Brief testing was done to see if removing channel 7 from the probe-channel pool would allow the probe's signal to be seen. However, it appears that the analog properties of both the NUC-140 board and the probe itself cause the input sine wave to become a sawtooth wave. This limited the system to using 1:1 (essentially directly wired) probes.

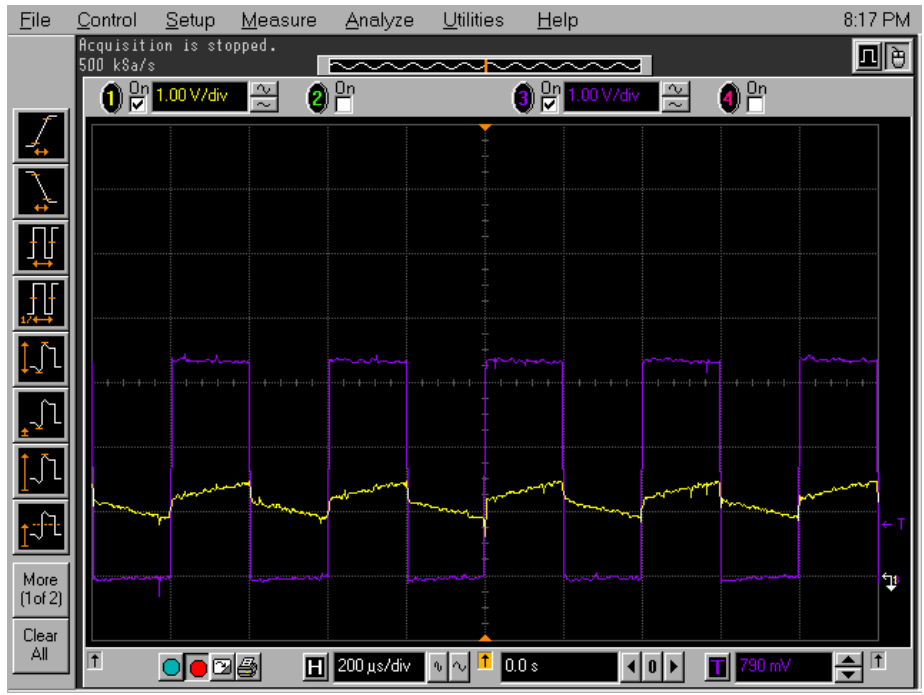

The yellow signal should be a scaled version of the purple calibration signal. However, due to the effects of the probe and the NUC-140 board, this signal has become distorted.

## Discussion and Conclusion

After completing development and verifying the system's performance, the resulting system became a very competent debugging tool. It implements 90% of the features I use on my original oscilloscope, being:
- Automatic, edge-triggered, and single modes.
- Waveform visualization using vertical and horizontal scaling.
- Probe Calibration.

Since most of my projects are low speed (topping out at a maximum of 300 Hz), this system is more than adequate. Trading off speed for portability makes complete sense, given that the NUC-140 oscilloscope is orders of magnitude smaller, lighter, and portable. I could easily see a more fully-fledged version of this system sharing a home with my other portable tools, such as my Ohm Meter.

The one feature that did not make it into the final product was the use of non 1:1 scope probes. Unfortunately, the reason that these scope probes won't work stems from design choices on the NUC-140 development board side. In terms of future work, creating a custom board that includes both the NUC-140 IC and the BNC connections would allow for better control over these factors. This would also allow for shrinking the total size down and properly encasing the system.

There is also room for improvement in increasing the potential speed of the ADC. While reading more on the ADC in the NUC-140's technical reference manual, it appears that there is a feature that allows the system to asynchronously copy data out of the ADC result registers directly into RAM. Not only that, but the ADC system itself can trigger this copy. Using this DMA system to perform the ADC to memory transfer instead of using the CPU could be used to increase the bandwidth of the scope.

Even with the current limitations, the system has shown great progress. It has been able to measure and display a wide range of signals present in real systems. With a bit more design work and tighter integration with the NUC-140 IC, it is possible to further improve the system. Overall, the project has been a great success and a valuable tool in working with larger-scale embedded projects.

This work is inspired by the embedded systems and digital design research group at UCCS. This group has done extensive work in embedded hardware and software architectures, techniques, and associated models. Their analyses [3],[4] shows that FPGA-based embedded systems are currently the best option to support applications and techniques, such as the ones presented in this report. Also, their previous work on FPGA-based embedded accelerators, architectures, and techniques for various compute and data-intensive applications, including data analytics/mining [5],[6],[7],[8],[9],[10],[11],[12],[13],[14]; control systems [15],[16],[17],[18],[19],[20]; cybersecurity [21],[22],[23]; machine learning [24],[25],[26],[27],[28],[29],[30]; communications [31],[32]; edge computing [33],[34],[35]; bioinformatics [36],[37]; and neuromorphic computing [38],[39]; demonstrated that FPGA-based embedded systems are the best avenue to support and accelerate complex algorithms and techniques.

Also as future work, we are planning to investigate FPGA-based hardware optimization techniques, such as parallel processing architectures (similar to [26],[40],[41],[42]), partial and dynamic reconfiguration traits (as stated in [43],[44],[45]) and architectures (similar to [22],[46],[47],[48]), HDL code optimization techniques (as stated in [49],[50]), and multi-ported memory architectures (similar to [51],[52],[53],[54]), to further enhance the performance metrics of FPGA-based embedded architectures for mini oscilloscope, while considering the associated tradeoffs.

# References

[1] Sitronix ST7565R. (2006, January 26). https://www.seacomp.com/sites/default/files/references/Sitronix-ST7565R.pdf
[2] Nuvoton. (2013, April 4). NumicroTM NUC130/Nuc140 Technical Reference Manual. https://www.nuvoton.com/resource-files/TRM_NUC130_NUC140(CN)_Series_EN_Rev2.05.pdf

[3] D.G. Perera and K.F. Li, "Analysis of Single-Chip Hardware Support for Mobile and Embedded Applications," in Proc. of IEEE Pacific Rim Int. Conf. on Communication, Computers, and Signal Processing, (PacRim'13), pp. 369-376, Victoria, BC, Canada, August 2013.

[4] D.G. Perera and K.F. Li, "Analysis of Computation Models and Application Characteristics Suitable for Reconfigurable FPGAs", in Proc. of 10th IEEE Int. Conf. on P2P, Parallel, Grid, Cloud, and Internet Computing, (3PGCIC'15), pp. 244-247, Krakow, Poland, Nov. 2015.

[5] S.N. Shahrouzi and D.G. Perera, "Optimized Hardware Accelerators for Data Mining Applications on Embedded Platform: Case Study Principal Component Analysis," Elsevier Journal on Microprocessor and Microsystems (MICPRO), vol. 65, pp. 79-96, March 2019.

[6] D.G. Perera and K.F. Li, "Embedded Hardware Solution for Principal Component Analysis," in Proc. of IEEE Pacific Rim Int. Conf. on Communication, Computers, and Signal Processing, (PacRim'11), pp.730-735, Victoria, BC, Canada, August 2011.

[7] D.G. Perera and Kin F. Li, "Hardware Acceleration for Similarity Computations of Feature Vectors," IEEE Canadian Journal of Electrical and Computer Engineering, (CJECE), vol. 33, no. 1, pp. 21-30, Winter 2008.

[8] D.G. Perera and K.F. Li, "On-Chip Hardware Support for Similarity Measures," in Proc. of IEEE Pacific Rim Int. Conf. on Communication, Computers, and Signal Processing, (PacRim'07), pp. 354-358, Victoria, BC, Canada, August 2007.

[9] K.F. Li and D.G. Perera, "An Investigation of Chip-Level Hardware Support for Web Mining," in Proc. of IEEE Int. Symp. on Data Mining and Information Retrieval, (DMIR'07), pp. 341-348, Niagara Falls, ON, Canada, May 2007.

[10] K.F. Li and D.G. Perera, "A Hardware Collective Intelligent Agent", Transactions on Computational Collective Intelligence, LNCS 7776, Springer, pp. 45-59, 2013.

[11] J.R. Graf and D.G. Perera, "Optimizing Density-Based Ant Colony Stream Clustering Using FPGA-Based Hardware Accelerator", in Proc. Of IEEE Int. Symp. on Circuits and Systems (ISCAS'23), 5-page manuscript, Monterey, California, May 2023.

[12] D.G. Perera, "Chip-Level and Reconfigurable Hardware for Data Mining Applications," PhD Dissertation, Department of Electrical & Computer Engineering, University of Victoria, Victoria, BC, Canada, April 2012.

[13] S. Navid Shahrouzi, "Optimized Embedded and Reconfigurable Hardware Architectures and Techniques for Data Mining Applications on Mobile Devices", PhD Dissertation, Department of Electrical & Computer Engineering, University of Colorado Colorado Springs, December 2018.

[14] J. Graf, "Optimizing Density-Based Ant Colony Stream Clustering Using FPGAs", MSc Thesis, Department of Electrical & Computer Engineering, University of Colorado Colorado Springs, CO, USA, March 2022.

[15] A.K. Madsen and D.G. Perera, "Efficient Embedded Architectures for Model Predictive Controller for Battery Cell Management in Electric Vehicles", EURASIP Journal on Embedded Systems, SpringerOpen, vol. 2018, article no. 2, 36-page manuscript, July 2018.

[16] A.K. Madsen, M.S. Trimboli, and D.G. Perera, "An Optimized FPGA-Based Hardware Accelerator for Physics-Based EKF for Battery Cell Management", in Proc. of IEEE Int,l Symp, on Circuits and Systems, (ISCAS'20), 5-page manuscript, Seville, Spain, May 2020.

[17] A.K. Madsen and D.G. Perera, "Towards Composing Efficient FPGA-Based Hardware Accelerators for Physics-Based Model Predictive Control Smart Sensor for HEV Battery Cell Management", IEEE ACCESS, (Open Access Journal in IEEE), pp. 106141-106171, 25th September 2023.

[18] A.K. Madsen and D.G. Perera, "Composing Optimized Embedded Software Architectures for Physics-Based EKF-MPC Smart Sensor for Li-Ion Battery Cell Management", Sensors, MDPI open access journal, Intelligent Sensors Section, 21-page manuscript, vol. 22, no. 17, 26th August 2022.

[19] A.K. Madsen, "Optimized Embedded Architectures for Model Predictive Control Algorithms for Battery Cell Management Systems in Electric Vehicles"; PhD Dissertation, Department of Electrical & Computer Engineering, University of Colorado Colorado Springs, August 2020.

[20] D. Abillar, "An FPGA-Based Linear Kalmann Filter for a Two-Phase Buck Converter Application", MSc Thesis, Department of Electrical & Computer Engineering, University of Colorado Colorado Springs, CO, USA, April 2024.

[21] A. Alkamil and D.G. Perera, "Efficient FPGA-Based Reconfigurable Accelerators for SIMON Cryptographic Algorithm on Embedded Platforms", in Proceedings of the IEEE International Conferences on Reconfigurable Computing and FPGAs, (ReConFig'19), 8-page manuscript, Cancun, Mexico, December 2019.

[22] A. Alkamil and D.G. Perera, "Towards Dynamic and Partial Reconfigurable Hardware Architectures for Cryptographic Algorithms on Embedded Devices", IEEE Access, Open Access Journal in IEEE, vol. 8, pp: 221720 – 221742, 10th December 2020.

[23] A. Alkamil, "Dynamic Reconfigurable Architectures to Improve Performance and Scalability of Cryptosystems on Embedded Systems", PhD Dissertation, Department of Electrical & Computer Engineering, University of Colorado Colorado Springs, 5th February 2021.

[24] M.A. Mohsin and D.G. Perera, "An FPGA-Based Hardware Accelerator for K-Nearest Neighbor Classification for Machine Learning on Mobile Devices", in Proceedings of the IEEE/ACM International Symposium on Highly Efficient Accelerators and Reconfigurable Technologies, (HEART'18), 6-page manuscript, Toronto, Canada, June 2018.

[25] S. Ramadurgam and D.G. Perera, "An Efficient FPGA-Based Hardware Accelerator for Convex Optimization-Based SVM Classifier for Machine Learning on Embedded Platforms", Electronics, MDPI open access journal, 36-page manuscript, vol. 10, no. 11, 31st May 2021.

[26] S. Ramadurgam and D.G. Perera, "A Systolic Array Architecture for SVM Classifier for Machine Learning on Embedded Devices", in Proc. of IEEE Int. Symp. on Circuits and Systems (ISCAS'23), 5-page manuscript, Monterey, California, May 2023.

[27] Jordi P. Miró, Mokhles A. Mohsin, Arkan Alkamil and Darshika G. Perera, "FPGA-based Hardware Accelerator for Bottleneck Residual Blocks of MobileNetV2 Convolutional Neural Networks", in Proceedings of the IEEE Mid-West Symposium on Circuits and Systems (MWCAS'25), 5-page manuscript, Lansing MI, August 2025.

[28] M. A. Mohsin, "An FPGA-Based Hardware Accelerator for K-Nearest Neighbor Classification for Machine Learning", MSc Thesis, Department of Electrical & Computer Engineering, University of Colorado Colorado Springs, CO, USA, December 2017.

[29] S. Ramadurgam, "Optimized Embedded Architectures and Techniques for Machine Learning Algorithms for On-Chip AI Acceleration", PhD Dissertation, Department of Electrical & Computer Engineering, University of Colorado Colorado Springs, 12th February 2021.

[30] J. P. Miro, " FPGA-Based Accelerators for Convolutional Neural Networks on Embedded Devices", MSc Thesis, Department of Electrical & Computer Engineering, University of Colorado Colorado Springs, CO, USA, May 2020.

[31] J. Nurmi and D.G. Perera, "Intelligent Cognitive Radio Architecture Applying Machine Learning and Reconfigurability" in Proc. of IEEE Nordic Circuits and Systems (NorCAS'21) Conf., 6-page manuscript, Oslo, Norway, October 2021.

[32] Kevin Young and Darsika G. Perera, "High-Level Synthesis Based FPGA Accelerator for GPS Signal Image Feature Extraction", in Proceedings of the IEEE Mid-West Symposium on Circuits and Systems (MWCAS'25), 5-page manuscript, Lansing MI, August 2025.

[33] D.G. Perera, "Reconfigurable Architectures for Data Analytics on Next-Generation Edge-Computing Platforms", Featured Article, IEEE Canadian Review, vol. 33, no. 1, Spring 2021. DOI: 10.1109/MICR.2021.3057144.

[34] M.A. Mohsin, S.N. Shahrouzi, and D.G. Perera, "Composing Efficient Computational Models for Real-Time Processing on Next-Generation Edge-Computing Platforms" IEEE ACCESS, (Open Access Journal in IEEE), 30-page manuscript, 13th February 2024.

[35] Mokhles A. Mohsin, and Darshika G. Perera, "Composing Efficient Computational Models for Real-Time Processing on Next-Generation Edge-Computing Platforms", IEEE ACCESS, (Open Access Journal in IEEE), 24-page manuscript, December 2025. DOI:10.1109/ACCESS.2025.3645763, (Impact factor 3.9).

[36] Laura H. Garcia, Arkan Alkamil, Mokhles A. Mohsin, Johannes Menzel and Darshika G. Perera, "FPGA-Based Hardware Architecture for Sequence Alignment by Genetic Algorithm", in Proceedings of the IEEE International Symposium on Circuits and Systems (ISCAS'25), 5-page manuscript, London, UK, May 2025.

[37] L. H. Garcia, "An FPGA-Based Hardware Accelerator for Sequence Alignment by Genetic Algorithm", MSc Thesis, Department of Electrical & Computer Engineering, University of Colorado Colorado Springs, CO, USA, December 2019.

[38] R.K. Chunduri and D.G. Perera, "Neuromorphic Sentiment Analysis Using Spiking Neural Networks", Sensors, MDPI open access journal, Sensing and Imaging Section, 24-page manuscript, vol. 23, no. 7701, 6th September 2023.

[39] S. Sharma and D. G. Perera, "Analysis of Generalized Hebbian Learning Algorithm for Neuromorphic Hardware Using Spinnaker" 8-page manuscript, https://doi.org/10.48550/arXiv.2411.11575

[40] R. Raghavan and D.G. Perera, "A Fast and Scalable FPGA-Based Parallel Processing Architecture for K-Means Clustering for Big Data Analysis", in Proceedings of the IEEE Pacific Rim International Conference on Communications, Computers, and Signal Processing, (PacRim'17), pp. 1-8, Victoria, BC, Canada, August 2017.

[41] D.G. Perera and Kin F. Li, "Parallel Computation of Similarity Measures Using an FPGA-Based Processor Array," in Proceedings of 22nd IEEE International Conference on Advanced Information Networking and Applications, (AINA'08), pp. 955-962, Okinawa, Japan, March 2008.

[42] R. Raghavan, "A Fast and Scalable Hardware Architecture for K-Means Clustering for Big Data Analysis", MSc Thesis, (Supervisor Dr. Darshika G. Perera), Department of Electrical & Computer Engineering, University of Colorado Colorado Springs, CO, USA, May 2016.

[43] D.G. Perera and K.F. Li, "A Design Methodology for Mobile and Embedded Applications on FPGA-Based Dynamic Reconfigurable Hardware", International Journal of Embedded Systems, (IJES), Inderscience publishers, 23-page manuscript, vol. 11, no. 5, Sept. 2019.

[44] D.G. Perera, "Analysis of FPGA-Based Reconfiguration Methods for Mobile and Embedded Applications", in Proceedings of 12th ACM FPGAWorld International Conference, (FPGAWorld'15), pp. 15-20, Stockholm, Sweden, September 2015.

[45] D.G. Perera and K.F. Li, "Discrepancy in Execution Time: Static Vs. Dynamic Reconfigurable Hardware", IEEE Pacific Rim Int. Conf. on Communications, Computers, and Signal Processing, (PacRim'24), 6-page manuscript, Victoria, BC, Canada, August 2024.

[46] D.G. Perera and Kin F. Li, "FPGA-Based Reconfigurable Hardware for Compute Intensive Data Mining Applications", in Proc. of 6th IEEE Int. Conf. on P2P, Parallel, Grid, Cloud, and Internet Computing, (3PGCIC'11), pp. 100-108, Barcelona, Spain, October 2011.

[47] D.G. Perera and Kin F. Li, "Similarity Computation Using Reconfigurable Embedded Hardware," in Proceedings of 8th IEEE International Conference on Dependable, Autonomic, and Secure Computing (DASC'09), pp. 323-329, Chengdu, China, December 2009.

[48] S.N. Shahrouzi and D.G. Perera, "Dynamic Partial Reconfigurable Hardware Architecture for Principal Component Analysis on Mobile and Embedded Devices", EURASIP Journal on Embedded Systems, SpringerOpen, vol. 2017, article no. 25, 18-page manuscript, 21st February 2017.

[49] S.N Shahrouzi and D.G. Perera, "HDL Code Optimization: Impact on Hardware Implementations and CAD Tools", in Proc. of IEEE Pacific Rim Int. Conf. on Communications, Computers, and Signal Processing, (PacRim'19), 9-page manuscript, Victoria, BC, Canada, August 2019.

[50] I.D. Atwell and D.G. Perera, "HDL Code Variation: Impact on FPGA Performance Metrics and CAD Tools", IEEE Pacific Rim Int. Conf. on Communications, Computers, and Signal Processing, (PacRim'24), 6-page manuscript, Victoria, BC, Canada, August 2024.

[51] S.N. Shahrouzi, A. Alkamil, and D.G. Perera, "Towards Composing Optimized Bi-Directional Multi-Ported Memories for Next-Generation FPGAs", IEEE Access, Open Access Journal in IEEE, vol. 8, no. 1, pp. 91531-91545, 14th May 2020.

[52] S.N. Shahrouzi and D.G. Perera, "An Efficient Embedded Multi-Ported Memory Architecture for Next-Generation FPGAs", in Proceedings of 28th Annual IEEE International Conferences on Application-Specific Systems, Architectures, and Processors, (ASAP'17), pp. 83-90, Seattle, WA, USA, July 2017.

[53] S.N. Shahrouzi and D.G. Perera, "An Efficient FPGA-Based Memory Architecture for Compute-Intensive Applications on Embedded Devices", in Proceedings of the IEEE Pacific Rim International Conference on Communications, Computers, and Signal Processing, (PacRim'17), pp. 1-8, Victoria, BC, Canada, August 2017.

[54] S.N. Shahrouzi and D.G. Perera, "Optimized Counter-Based Multi-Ported Memory Architectures for Next-Generation FPGAs", in Proceedings of the 31st IEEE International Systems-On-Chip Conference, (SOCC'18), pp. 106-111, Arlington, VA, Sep. 2018.